\documentstyle[twoside,fleqn,espcrc2]{article}

\newcommand{\AmS}{{\protect\the\textfont2
  A\kern-.1667em\lower.5ex\hbox{M}\kern-.125emS}}

\title{Lattice charge overlap and the elastic limit
 \author{Walter Wilcox \\ Department of Physics,
        Baylor University, Waco, TX 76798, USA}}

\begin{document}

\begin{abstract}
The results of a lattice simulation of
time-separated charge overlap for the
charged pion are discussed. The expected result
$\sim {\rm exp}[-(E_q - m_{\pi})t\,]$ for large charge density
overlap time separations, $t$, is
clearly visible in the Fourier transform, indicating that the elastic
limit can be achieved at low to medium momentum values
on present-sized lattices.
The implications of this result for
direct lattice simulations of hadron structure functions are discussed
and a brief presentation of the lattice formalism is given.
\end{abstract}

\maketitle

\section{INTRODUCTION}

In trying to better understand the internal structure of
hadrons in lattice QCD, charge overlap techniques
are playing an increasingly important role. These methods
are based upon the
simulation of the hadronic matrix elements
$<h({\bf 0})|T[J^{d}_{\mu}({\bf r},t)J^{u}_{\nu}(0)]|h({\bf 0})>$,
where $J^{d,u}_{\mu}$ are $d,u$ flavor current densities.
In their original form\cite{barad,me1} these measurements,
using $t=0$ between the currents, allowed qualitative
studies of lattice hadrons to be carried out.
However, it has been pointed out that large Euclidean
time separations of the currents allows one to
extract form factor data from this matrix element
in the so-called elastic limit\cite{me2}.
This is interesting in it's own right, but
it is also clear that by replacing the flavor currents
in this matrix element with the full electromagnetic
ones, we are then studying the basic
matrix element needed for hadron structure
functions\cite{me4}. Thus, in moving from the
qualitative to the quantitative stage,
the elastic limit of these matrix elements forms an
important bridge, leading to
both elastic and inelastic properties
of hadrons. The question of how readily
this limit may be implemented in lattice simulations is the
subject of the present investigation.

\section{THEORY}

Consider the lattice Euclidean
time-separated charge overlap
distribution for zero momentum pions:
\begin{equation}
{\cal Q}_{00}^{du}({\bf q}^{2},t) \equiv
\sum_{{\bf r}}
e^{-i{\bf q}\cdot{\bf r}}{\cal P}_{00}^{du}({\bf r},t),\label{1}
\end{equation}
where
\begin{eqnarray}
\lefteqn{{\cal P}_{00}^{du}({\bf r},t)  \equiv}  \nonumber\\
\lefteqn{\sum_{{\bf x}}<\pi^{+}({\bf 0})|
T[-\rho^{d}({\bf r}+{\bf x},t)
\rho^{u}({\bf x},0)]|\pi^{+}({\bf 0})>.} \nonumber\\
\label{2}
\end{eqnarray}
$\rho^{u,d}$ are the $u$, $d$ flavor charge densities
and \lq $T$\,\rq is time-ordering.
Figure 1 is a symbolic representation of the
measurement in Eq.~({\ref{2}).
Assuming $t>0$ and
inserting a complete set of states, this
results in
\begin{eqnarray}
\lefteqn{{\cal Q}_{00}^{du}({\bf q}^{2},t) =
\sum_{X}<\pi^{+}({\bf 0})|-\rho^{d}(0)|X({\bf q})>} \nonumber\\
\lefteqn{\qquad\quad\cdot <X({\bf q})|\rho^{u}(0)|\pi^{+}({\bf 0})>
e^{-(E_{X}-m_{\pi})t}.} \label{3}
\end{eqnarray}
For large Euclidean times the sum reduces to a
single term and we find in the continuum limit that
\begin{eqnarray}
\lefteqn{{\cal Q}_{00}^{du}({\bf q}^{2},t)
\stackrel{t\,\gg 1}{\longrightarrow}} \nonumber \\
\lefteqn{\qquad\qquad\quad\frac{(E_{q}+m_{\pi})^{2}}
{4E_{q}m_{\pi}}F_{\pi}^{2}(q^{2})\,e^{-(E_{q}-m_{\pi})t},} \label{4}
\end{eqnarray}
when the SU(2) flavor symmetry is present.

\begin{figure}[htb]

\caption{Symbolic representation of the $d,u$ charge
overlap measurement.}
\label{figure1}
\end{figure}

Thus, by separating the currents in time
it is in principle possible to damp out the intermediate
state contributions and to measure form factors
at arbitrary lattice momenta. On a finite space-time lattice,
however, achieving this elastic limit is problematical. The exponential
damping factor, $e^{-(E_{q}-m_{\pi})t}$, is not large and the
fixed time positions of the initial and final pion interpolating
fields limit the possible charge density separation times.
We now turn to a numerical investigation of this limit.

\section{SIMULATION}

The present discussion is based on the
numerical results found in Ref.~\cite{me4},
where more calculational details may be found.
This study was conducted on $12$
quenched $\beta=6.0$ SU(3) configurations
($16^{3}\times 24$) at $\kappa=.154$ (the
largest $\kappa$ value studied in
Ref.~\cite{nucleon}.)
The pion interpolating fields were located at
time slices $4$ and $21$ and the
current densities
were positioned as symmetrically possible
in time between these sources.
Two Wilson quark propagators per configuration,
with origins at the positions of the interpolating sources,
were necessary to extract the relative overlap function
${\cal P}^{du}_{00}({\bf r}, t)$ in Eq.~(\ref{2}).
(${\cal P}^{du}_{00}({\bf r}, t)$ is actually the large
Eucldean time limit of a similarly denoted quantity in
Ref.~\cite{me4}.)
The pion interpolating field at time slice $21$
was constructed with quark propagators which
were smeared over the entire spatial
volume using the lattice Coulomb gauge.
Both point and smeared pion fields were
used at time slice 4. These propagators
produce four-point functions which project exactly on zero
pion momentum. Charge density self-contractions were
neglected in forming these quantities.

\begin{figure}[htb]
\vspace{65mm}
\caption{${\rm Log}_{10}$ plot of the Fourier transform,
${\cal Q}^{du}_{00}$, of the
spatial density charge correlation function as
a function of relative time separation, $t$, between charge
density operators.}
\label{figure2}
\end{figure}

Fig.~2 represents a ${\rm log}_{10}$ plot of
the Fourier transform of these distribution
functions, ${\cal Q}_{00}^{du}({\bf q}^{2},t)$,
at the two lowest lattice spatial momentum values, $|{\bf q}|=\pi/8$
(upper points), $\sqrt{2}(\pi/8)$
(lower points), as a function of relative time separation between
current densities. The solid lines shown in this figure come from
the expected asymptotic exponential falloff specified by Eq.~(\ref{4}),
using the (smeared) $\kappa=.154$
data from Table 1 ($m_{\pi}=.369, m_{\rho}=.46$) of
Ref.~\cite{nucleon}, assuming the vector dominance form for the pion
form factor: $F_{\pi}(q^{2})=1/(1+q^{2}/m_{\rho}^{2})$.
(We also assume the continuum relation $E_{q}=(m_{\pi}^{2}
+ {\bf q\,}^{2})^{1/2}$.) Actually shown
in this figure are results for both point-to-smeared ($\Diamond$)
as well as smeared-to-smeared ($\Box$) correlation functions.
In all cases, the expected functional dependence $\sim
e^{-(E_{q}-m_{\pi})t}$ is present, indicating that by
time step $7$ or $8$ single exponential behavior has emerged.
This is similar to the number of time steps needed in
hadron spectrum calculations.
This behavior is remarkable because although
we are damping out intermediate states as $t$ increases,
we are also moving closer to
possible contaminations from the
fixed interpolating fields at either
time end. In fact, we do not see any indications
of such contamination in the data.
These correlation functions are also unusual because the asymptotic
approach is from below, indicating damping of
negative terms in the $d, u$ correlation function.
\begin{figure}[htb]
\vspace{65mm}
\caption{Extracted pion form factor, $F_{\pi}$, as a
function of $q^{2}/m_{\rho}^{2}$. The solid line is vector
dominance.}
\label{figure3}
\end{figure}
Ref.~\cite{me5} indicates that in the continuum limit,
these are primarily positive G-parity states\cite{foot1}.

Although the point-to-smeared and smeared-to-smeared
results exhibit essentially overlapping
error bars, the smeared-to-smeared values are
systematically low compared to the
point-to-smeared results, indicating a slight dependence on the
form of the interpolating field used. However, the results
of Ref.~\cite{me2} indicate that such a dependence decreases as
$\kappa\rightarrow\kappa_{cr}$, that is, as the physical
regime is approached.

By fitting time steps 7 through 10 of Fig.~2 with a
single exponential of known slope and removing the
kinematical factor in Eq.~(\ref{4}), a value of
$F_{\pi}(q^{2})$ can be obtained. The results for the two
calculated $q^{2}/m_{\rho}^{2}$ values using point-to-smeared
correlation functions are shown in Fig.~3 ($\Diamond$).
The values found are consistent within errors with vector
dominance, shown as the solid line. Also shown in this figure
are the results ($\Box$) from a previous three-point-function simulation
of the pion form factor\cite{pion}. Comparison
shows that the error bars of these different
techniques are of the same order of magnitude for similar numbers
of configurations. (Note the different $\beta$
values and lattice sizes of the two simulations, however.)

\section{STRUCTURE FUNCTIONS}

In order to begin to understand the significance of these
results for structure functions, let us recall the basic form of
these quantities (continuum Minkowski expression):
\begin{eqnarray}
\lefteqn{W_{\alpha\beta}(q^{2},\nu)= \frac{1}{2\pi}\int d^{4}x\,
 e^{-iq\cdot x}} \nonumber\\
\lefteqn{\qquad\qquad\cdot\,\frac{1}{2m}(\pi^{+} ({\bf 0})|
[J_{\alpha}(x),J_{\beta}(0)]|\pi^{+}({\bf 0}))\, .}
\label{5}
\end{eqnarray}
We adopt the lab frame
where the external pion momentums are zero and use the
full electromagnetic currents.
In this frame $\nu \equiv -p\cdot q/m =
q_{0}$. In order to make contact
with lattice expressions, we imagine discretizing space, but
keeping the time variable continuous. Using the
correspondences,
\begin{eqnarray}
\qquad\int d^{3}x & \rightarrow & a^{3}
\displaystyle{\sum_{{\bf x}}\, ,} \nonumber\\
\qquad |\pi^{+}({\bf p})) & \rightarrow &
[N_{s}a^{3}2E_{p}]^{\frac{1}{2}}|\pi^{+}({\bf p})>, \nonumber\\
\qquad J_{\alpha}^{cont.} & \rightarrow & a^{-3}
J_{\alpha}^{latt.}, \nonumber
\end{eqnarray}
where $N_{s}$ is the number of space points in the lattice,
this gives (all quantities on the lattice)
\begin{eqnarray}
\lefteqn{W_{\alpha\beta}(q^{2},\nu)= \frac{N_{s}}{2\pi}\int_{-\infty}
^{\infty} dt\, e^{i\nu t}\sum_{{\bf x}}
e^{-i{\bf q}\cdot{\bf x}}} \nonumber\\
\lefteqn{\qquad\qquad\cdot<\pi^{+} ({\bf 0})|
[J_{\alpha}(x),J_{\beta}(0)]|\pi^{+}({\bf 0})>.}
\label{6}
\end{eqnarray}
By a standard set of manipulations, this can be shown to
result in the expression:
\begin{eqnarray}
\lefteqn{W_{\alpha\beta}(q^{2},\nu)= N_{s}^{2}
\displaystyle{\sum_{X}}\delta(\nu - E_{X} + m_{\pi})} \nonumber\\
\lefteqn{\cdot<\pi^{+} ({\bf 0})|
J_{\alpha}(0)|X({\bf q})>
<X({\bf q})|J_{\beta}(0)|\pi^{+}({\bf 0})>.} \nonumber\\
\label{7}
\end{eqnarray}
How is this quantity to be measured on the lattice?
Let us consider the
generalization of Eqs.~(\ref{1}) and (\ref{2}) to
arbitrary components of the full electromagnetic current.
Define in Euclidean space
\begin{eqnarray}
\lefteqn{{\cal Q}_{\alpha\beta}({\bf q}^{2},t)\equiv
\displaystyle{\sum_{{\bf r},{\bf x}}}
e^{-i{\bf q}\cdot{\bf r}}} \nonumber\\
\lefteqn{\cdot<\pi^{+} ({\bf 0})|
T[J_{\alpha}({\bf x}+{\bf r},t)
J_{\beta}({\bf x},0)]|\pi^{+}({\bf 0})>.}
\label{8}
\end{eqnarray}
Again assuming $t>0$, this results in
\begin{eqnarray}
\lefteqn{{\cal Q}_{\alpha\beta}({\bf q}^{2},t)=N_{s}^{2}
\displaystyle{\sum_{X}}
e^{-(E_{X}-m_{\pi})t}} \nonumber\\
\lefteqn{\cdot<\pi^{+} ({\bf 0})|
J_{\alpha}(0)|X({\bf q})>
<X({\bf q})|J_{\beta}(0)|\pi^{+}({\bf 0})>.} \nonumber\\
\label{9}
\end{eqnarray}
We notice that the connection between
Eq.~(\ref{9}) and (\ref{7}) is formally given by the
inverse Laplace transform,
\begin{eqnarray}
W_{\alpha\beta}(q^{2},\nu)=\frac{1}{2\pi i}
\int_{c-i\infty}^{c+i\infty} dt\,
e^{\nu t}{\cal Q}_{\alpha\beta}({\bf q}^{2},t)\, ,
\label{11}
\end{eqnarray}
where $c > 0$ and the contour is closed in
the left-half complex t-plane.
Since we are at fixed spatial momentum
${\bf q}$, the lattice data extracted will actually cut
a parabola-like path given
by $x=q^{2}/(2m\sqrt{{\bf q}^{2}-q^{2}})$ in the
$x$, $q^{2}$ plane
($x \equiv q^{2}/(2m\nu)$ and
$q^{2}\equiv {\bf q}^{2}-q_{0}^{2}$).
In the lab frame the range of $\nu$ when
$0<x<1$ is $|{\bf q}| < \nu <(E_{q}-m_{\pi})$.
One can assure these
kinematical constraints in the lattice model by restricting
the assumed forms of ${\cal Q}_{\alpha\beta}({\bf q}^{2},t)$
to assure contributions to $W_{\alpha\beta}(q^{2},\nu)$
only within the continuum
range of $\nu$.
We also need to make a quasi-continuum assumption
about the lattice data: quantities like $W_{00}(q^{2},\nu)$
should be very poorly fit by a positively
weighted sum of exponentials.
This assumption implies that ${\cal Q}_{00}({\bf q}^{2},t)$
is given by products of exponentials and
inverse powers of $t$. (In this case Eq.~(\ref{7})
becomes purely formal but Eq.~(\ref{11}) continues
to hold.) This whole discussion
has been carried out for the pion, but it is clear that
there is no barrier to applying these
techniques to the phenomenologically more
interesting proton case as well.

\section{DISCUSSION}
There are now two workable techniques for
extracting form factor data from lattice simulations:
direct current insertion\cite{direct} and elastic
charge overlap. The elastic limit
also makes it possible to perform
direct simulations of hadron structure functions.
Attention so far has been focused on the moments of such functions,
which are given by the operator product expansion.
These expansions
are based upon separation of the short-distance physics,
calculated perturbatively, from the long-distance part,
which can be evaluated on the lattice
in the form of certain
operator expectation values. Because of the
perturbative assumption these methods work best at
large $q^{2}$; direct lattice simulations must
use low dimensionless $(qa)^{2}$, so these two techniques
should be complementary. The question of whether the
direct method can reach the scaling regime is still
open, but note that the chiral limit (elastic) $q^{2}$
range in Ref.~\cite{nucleon} was
$.6\leq q^{2}\leq 1.9 {\rm GeV}^{2}$.
Assuming the elastic limit of all necessary
flavor-diagonal and non-diagonal
four-point functions can be demonstrated,
direct simulations of structure functions should be
feasible with current computer technology.

This work was partially supported by the National Center for Supercomputing
Applications and the NSF
under Grant No.\ PHY-$9203306$
and utilized the NCSA CRAY~2 system at the University of
Illinois at Urbana-Champaign.

\end{document}